\documentclass[preprint,trackchanges]{aastex62}

\hypersetup{linkcolor=magenta, citecolor=cyan, filecolor=yellow, urlcolor=blue}

\received{2019 January 15}
\revised{2019 August 26}
\accepted{2019 September 1}
\published{2019 October 16}

\shorttitle{``Double-tracking'' Characteristic of the Spectral Evolution of GRB 131231A}
\shortauthors{Li et al.}

\begin{document}

\title{``Double-tracking'' Characteristic of the Spectral Evolution of GRB 131231A: Synchrotron Origin?}

\author[0000-0002-1343-3089]{Liang Li}
\affil{Purple Mountain Observatory, Chinese Academy of Sciences, Nanjing 210033, Peopleʼs Republic of China}
\affiliation{ICRANet, Piazza della Repubblica 10, I-65122 Pescara, Italy}

\correspondingauthor{Liang Li}
\email{liang.li@icranet.org, gengjinjun@nju.edu.cn, xfwu@pmo.ac.cn, yu.wang@inaf.it, z.lucas.uhm@gmail.com}

\author{Jin-Jun Geng}
\affiliation{School of Astronomy and Space Science, Nanjing University, Nanjing 210023, Peopleʼs Republic of China}
\affiliation{Key Laboratory of Modern Astronomy and Astrophysics (Nanjing University), Ministry of Education, Nanjing 210023, Peopleʼs Republic of China}

\author{Yan-Zhi Meng}
\affil{Purple Mountain Observatory, Chinese Academy of Sciences, Nanjing 210033, Peopleʼs Republic of China}
\affiliation{University of Chinese Academy of Sciences, Beijing 100049, Peopleʼs Republic of China}

\author{Xue-Feng Wu}
\affil{Purple Mountain Observatory, Chinese Academy of Sciences, Nanjing 210033, Peopleʼs Republic of China}

\author{Yong-Feng Huang}
\affiliation{School of Astronomy and Space Science, Nanjing University, Nanjing 210023, Peopleʼs Republic of China}
\affiliation{Key Laboratory of Modern Astronomy and Astrophysics (Nanjing University), Ministry of Education, Nanjing 210023, Peopleʼs Republic of China}

\author{Yu Wang}
\affiliation{ICRANet, Piazza della Repubblica 10, 65122 Pescara, Italy}
\affiliation{Dip. di Fisica and ICRA, Sapienza Universita di Roma, Piazzale Aldo Moro 5, I-00185 Rome, Italy}
\affiliation{INAF -- Osservatorio Astronomico d'Abruzzo, Via M. Maggini snc, I-64100, Teramo, Italy}

\author{Rahim Moradi}
\affiliation{ICRANet, Piazza della Repubblica 10, 65122 Pescara, Italy}
\affiliation{Dip. di Fisica and ICRA, Sapienza Universita di Roma, Piazzale Aldo Moro 5, I-00185 Rome, Italy}

\author{Z. Lucas Uhm}
\affiliation{Astrophysics Science Division, NASA Goddard Space Flight Center, Greenbelt, MD 20771, USA}
\affiliation{Korea Astronomy and Space Science Institute, Daejeon 34055, Republic of Korea}

\author{Bing Zhang}
\affiliation{Department of Physics and Astronomy, University of Nevada, Las Vegas, NV 89154, USA}

\begin{abstract}

The characteristics of the spectral evolution of the prompt emission of gamma-ray bursts (GRBs), which are closely related to the radiation mechanism (synchrotron or photosphere), are still an unsolved subject. Here, by performing the detailed time-resolved spectral fitting of GRB 131231A, which has a very bright and well-defined single pulse, some interesting spectral evolution features have been found.
(i) Both the low-energy spectral index $\alpha$ and the peak energy $E_{\rm p}$ exhibit the ``flux-tracking'' pattern (``double-tracking'' characteristics).
(ii) The parameter relations, i.e., $F$ (the energy flux)-$\alpha$, $F$-$E_{\rm p}$, and $E_{\rm p}$-$\alpha$, along with the analogous Yonetoku $E_{\rm p}$-$L_{\gamma,\rm iso}$ relation for the different time-resolved spectra, show strong monotonous (positive) correlations, both in the rising and the decaying phases.
(iii) The values of $\alpha$ do not exceed the synchrotron limit ($\alpha$= -2/3) in all slices across the pulse, favoring the synchrotron origin.
We argue that the one-zone synchrotron emission model with the emitter streaming away at a large distance from the central engine can explain all of these special spectral evolution characteristics. 

\end{abstract}


\section{Introduction} \label{sec:intro}

One of the leading models to interpret the observed spectral shape in the prompt emission of gamma-ray bursts (GRBs) is the synchrotron radiation model \cite[e.g.,][]{2000ApJ...543..722L, 2000AIPC..526..185T, 2004ApJ...613..460B, 2011ApJ...741...24B, 2014ApJ...784...17B, 2018pgrb.book.....Z}, which invokes emission of relativistic charged particles either from internal shocks or from internal magnetic dissipation processes.
The observed GRB spectra, i.e., both the time-integrated and the time-resolved spectra, can be described well by an empirical Band function \citep{1993ApJ...413..281B}---namely, the smoothly connected broken power law. The low-energy power-law index $\alpha $ is typically $\sim $ -1.0, the high-energy index $\beta $ $\sim $ -2.2, and the peak energy $E_{\mathrm{p}}$ $\sim $ 300 keV for the time-integrated spectrum, based on the statistical works of a large sample of GRBs (e.g., \citealt{2000ApJS..126...19P, 2006ApJS..166..298K, 2011A&A...531A..20G, 2011A&A...530A..21N, 2011ApJ...730..141Z, 2012ApJS..199...19G, 2013ApJ...764...75G}). For the time-resolved spectra, the low-energy index $\alpha $ is much harder ($\alpha \sim $ -0.8; \citealt{2006ApJS..166..298K, 2016A&A...588A.135Y, 2018arXiv181007313Y}). The high-energy spectral index $\beta$ is usually not evaluated for time-resolved spectra due to the small number of photons available.
The peak energy is, however, often different at the peak time from the average spectrum \citep{2006ApJS..166..298K}.

The evolution characteristics of $E_{\mathrm{p}}$ and $\alpha$ based on the time-resolved spectra have been widely studied in early (pre-\textit{Fermi} era; e.g., \citealt{1983Natur.306..451G, 1986ApJ...301..213N, 1994ApJ...426..604B, 1994ApJ...422..260K, 1995ApJ...439..307F, 1997ApJ...479L..39C, 2006ApJS..166..298K, 2009ApJ...698..417P}) and recent (\textit{Fermi} era; e.g., \citealt{2012ApJ...756..112L, 2016A&A...588A.135Y, 2017ApJ...846..137O, 2018MNRAS.475.1708A, 2018A&A...613A..16R, 2018arXiv181007313Y, 2019ApJS..242...16L, 2019ApJS..245...7L}) works. In the pre-\textit{Fermi} era, the $E_{\mathrm{p}}$ is revealed to exhibit several distinct patterns: (i) the ``hard-to-soft'' trend, decreasing monotonically regardless of the rise and fall of the flux \citep[e.g.,][]{1986ApJ...301..213N, 1994ApJ...426..604B, 1997ApJ...486..928B}; (ii) the ``flux-tracking'' trend \citep[e.g.,][]{1983Natur.306..451G, 1999ApJ...512..693R}; and (iii) others (e.g., soft-to-hard or chaotic evolutions; \citealt{1985ApJ...290..728L, 1994ApJ...422..260K}). After the launch of \textit{Fermi} in 2008, with the spectral data of higher quality, the former two patterns are confirmed to be dominated: ``hard-to-soft'' for about two-thirds and ``flux-tracking'' for about one-third \citep[e.g.,][]{2012ApJ...756..112L, 2018arXiv181007313Y}. The physical origin of these $E_{\mathrm{p}}$ evolution patterns still remain unsolved, though some scenarios have been proposed in the literature \citep[e.g.,][]{1997ApJ...479L..35L, 1999ApJ...512..693R, 2006ApJ...637..869M, 2011CRPhy..12..206Z, 2011ApJ...726...90Z, 2014ApJ...785..112D, 2014NatPh..10..351U, 2016ApJ...825...97U, 2018A&A...616A.138O, 2019A&A...628A..59O, 2018ApJ...869..100U, 2019A&A...627A.105B}.

As for the $\alpha $ evolution, based on a Burst And Transient Source Experiment sample, for the first time,  \cite{1997ApJ...479L..39C} pointed out that $\alpha $ evolves with time rather than remaining constant. Compared with $E_{\mathrm{p}}$ evolution, the $\alpha $ evolution is more chaotic, and thus there are relatively fewer studies and physical explanations. In addition, the correlation analysis for the evolution of $E_{\mathrm{p}}$ and $\alpha $ in a single burst is lacking. Here in this work, after carrying out the detailed time-resolved spectral analysis of the single pulse in the bright \textit{Fermi} burst, GRB 131231A, we find that both the $E_{\mathrm{p}}$ and $\alpha$ evolutions exhibit the ``flux-tracking'' behavior, which can be defined as ``double-tracking'' patterns of the spectral evolution. This is quite interesting, since such features are very rarely observed within a single burst. 
The low-energy power-law photon index $\alpha$, as predicted by synchrotron model, has a limit value called the line of death (LOD; \citealt{1998ApJ...506L..23P})\footnote{Recent studies suggested that the LOD is not a hard limit for synchrotron radiation. \cite{zhang16} showed that instead of the Band function fits, one should apply physical synchrotron models with properly treatment of synchrotron cooling to fit the original data. \cite{2018arXiv181006965B} showed that with such an approach, many bursts with Band-function $\alpha$ beyond the LOD can be actually well fit by the synchrotron model. This suggests that the LOD is no longer a hard limit for synchrotron radiation. From theoretical aspects, introducing small pitch angle synchrotron radiation \citep{2002ApJ...565..182L} or pitch-angle distribution \citep{yang18} can also help to break the LOD limit.}. This limit requires that $\alpha$ could not exceed the value of -2/3. On the other hand, when the electrons are in the fast cooling regime, the spectral index of the electron distribution is -2, resulting a photon index of -3/2 \citep{1998ApJ...497L..17S}. Therefore, in the simple synchrotron scenario, $\alpha$ ranges from -3/2 (fast cooling case) to -2/3 (slow cooling case).
Considering that $\alpha$ does not exceed the synchrotron limit ($\alpha $= -2/3; \citealt{1998ApJ...506L..23P}) in all slices across the pulse, we try to use the synchrotron emission model to interpret these ``double-tracking'' spectral evolution characteristics.

The paper is organized as follows. The data analysis is presented in Section 2. The physical interpretations are presented in Section 3. The conclusions and discussions are presented in Section 4. Throughout the paper, a concordant Friedmann-Lemaitre-Robertson-Walker Cosmology with parameters $H_{0}=71$ $\mathrm{km}$ $\mathrm{s^{-1}}$ $\mathrm{Mpc^{-1}}$, $\Omega _{M}=0.30$, and $\Omega _{\Lambda}=0.70$ are adopted. The convention $Q=10^{x}Q_{x}$ is adopted in cgs units.

\section{Data Analysis} \label{sec:data}

\subsection{Observations} 

GRB 131231A (trigger 410157919/131231198) triggered gamma-ray burst monitor (GBM: 8 keV - 40 MeV) on board the NASA {\it Fermi Gamma-Ray Observatory} at 04:45:16.08 UT ($T_{0}$) on 2013 December 31. 
In addition, the intense high-energy emission of GRB 131231A also triggered the Large Area Telescope (LAT) on board {\it Fermi}, and the Konus-{\it Wind}.
The light curve of the prompt emission exhibits a single large peak profile (Fig.\ref{fig:f1}), with $T_{90}$ \citep{1981Ap&SS..80....3M, 1993ApJ...413L.101K} of 31.23$\pm$0.57 s in the 50-300 keV band \citep{2014GCN.15644....1J}. 
GRB 131231A is a very bright burst, and the fluence in the energy range of 10 keV-1000 keV from $T_{0}$+0.003 s to $T_{0}$+56 s reported by the GBM team is (1.40$\pm$0.001) $\times$ 10$^{-4}$ erg cm$^{-2}$ \citep{2014GCN.15672....1J}, while in the energy range of 20 keV-10 MeV from $T_{0}$ to $T_{0}$+7.488, based on the observation of Konus-{\it Wind}, is (1.55$\pm$0.05) $\times$ 10$^{-4}$ erg cm$^{-2}$ \citep{2014GCN.15670....1G}.
The 1024 ms peak flux in the energy range of 10-1000 keV is 78.81 $\pm$ 0.65 photon cm$^{-2}$ s$^{-1}$ according to the {\it Fermi} observation. 
The time-averaged spectrum from $T_{0}$ s to $T_{0}$+34.303 s, as reported in \cite{2014GCN.15670....1G}, can be well fitted by the Band function \citep{1993ApJ...413..281B}, with the best-fit parameters of the low-energy photon index as $\alpha$=-1.28$\pm$0.04, the high-energy photon index as $\beta$=-2.47 $\pm$ 0.05 and the peak energy $E_{\rm p}$= 163$\pm$6 keV, and the value of fitting quality as $\chi^{2}$/dof = 94.3/82 \citep{2014GCN.15670....1G}. 
Furthermore, GeV afterglow emission was detected for GRB 131231A and its temporal and spectral behavior can be accounted for with the synchrotron self-Compton radiation of the relativistic electrons accelerated by the  forward shock \citep{2014ApJ...787L...6L}. {\it Swift}/Burst Alert Telescope (BAT) and the early {\it Swift}/X-ray Telescope (XRT) observations, as well as the optical afterglow emission, are not available.
The X-ray counterpart was detected by the {\it Swift}/XRT at 52.186 ks after the trigger, with the location of R.A.= 10.5904 and dec.= -1.6519 \citep{2014ApJ...787L...6L}.
The redshift of this GRB is $\sim$ 0.642 \citep{2014GCN.15652....1C, 2014GCN.15645....1X} and the estimation of the released isotropic energy is $E_{\gamma, \rm iso}$= (3.9$\pm$0.2)$\times$ 10$^{53}$ erg \citep{2014GCN.15645....1X}. 

\subsection{Time-resolved Spectral Fits} 

The software package RMFIT\footnote{\url{https://fermi.gsfc.nasa.gov/ssc/data/analysis/user/}} (version 3.3pr7) is applied to carry out the spectral analysis. To ensure consistency of the results across various fitting tools, we also compare the results with the Bayesian approach analysis package---namely, the Multi-Mission Maximum Likelihood Framework (3ML; \citealt{2015arXiv150708343V}), which has been applied to conduct the time-resolved spectral fitting analysis by many authors \citep[e.g.,][]{2017arXiv171008362B, 2018arXiv181007313Y,  2019ApJS..242...16L, 2019ApJS..245...7L, 2019arXiv190809240L}.
The GBM carries 12 sodium iodide (NaI, 8keV-1MeV) and two bismuth germanate (BG0, 200 keV-40 MeV) scintillation detectors \citep{2009ApJ...702..791M}.
We perform the spectral analysis using the data of three NaI detectors (n0, n3, and n4) and one BGO detector (b0) on {\it Fermi}-GBM. The Time-Tagged Events (TTE) data is used to contain pulse height counts and photon counts are, therefore, obtained after the true signal is deconvolved from the detector response. We estimate the background photon counts by fitting the light curve before and after the burst with a one-order background polynomial model. The source is selected as the interval from 0 to 50 s, which covers the main source interval after subtracting the background.
The time bin selection for the time-resolved spectral analysis follows the Bayesian Blocks method (BBlocks; \citealt{2013ApJ...764..167S}). We also calculate the signal-to-noise ratio (S/N) for each slice, with the derived photon signal and background noise using the XSPEC (version 12.9.0) tool\footnote{\url{https://heasarc.gsfc.nasa.gov/xanadu/xspec/}}.
To carry out a precise spectral analysis, enough source photons should be included in each slice. Therefore, a suitable value of S/N is required \citep{2018ApJS..236...17V}; we apply S/N $\geq$ 20 in this paper. 
After binning with the BBs, we obtain 26 spectra in the interval from 0 to 50 s. All of these spectra can be well fitted by the Band model, except for 3 spectra with unconstrained $\beta$. 
The goodness-of-fit is determined by reduced C-stat minimization. The best-fit parameters for each spectrum ($\alpha$, $\beta$ and $E_{\rm p}$), along with its time interval, S/N, C-stat/degrees of freedom (dof), and reduced C-stat, are summarized in Table \ref{tab:table1}. 
An example of count spectral fits is shown in Figure \ref{fig:f1}, and the temporal evolution of spectral parameters ($E_{\rm p}$ and $\alpha$) is presented in Figure \ref{fig:f2}.

\subsection{Parameter Correlation Analysis}

The spectral correlation analysis plays an important role in
revealing the radiation nature of GRB prompt emission. The key correlations include those between the energy flux $F$, the peak energy $E_{\rm p}$, and the low-energy photon index $\alpha$, i.e., $E_{\rm p}$-$F$, $\alpha$-$F$, and $E_{\rm p}$-$\alpha$ correlations.

To investigate the above mentioned relations, the energy flux $F$ in each slice needs to be known. We obtain the energy flux (erg cm$^{-2}$s$^{-1}$) by integrating the $F_{\rm E}$ (erg cm$^{-2}$s$^{-1}$keV$^{-1}$) spectrum of the Band model, for the energy range from 10 keV to 40 MeV, and the corresponding time interval of each time-resolved spectrum (Column (1) in Table 1). 
Then, we show the temporal evolution of $E_{\rm p}$ and $\alpha$, respectively, compared with the energy flux in Figure \ref{fig:f3}, and find the more prominent ``flux-tracking'' behavior than Figure \ref{fig:f3} (especially before the peak time).

The relation between the energy flux $F$ and $E_{\rm p}$, i.e., the Golenetskill $E_{\rm p}$-$F$ relation \citep{1983Natur.306..451G, 2019MNRAS.485.1262B}, for the time-resolved spectra of GRB 131231A is shown in Figure \ref{fig:f4}a. Previous analyses \cite[e.g.,][]{2001ApJ...548..770B, 2009MNRAS.393.1209F, 2010A&A...511A..43G, 2018arXiv181007313Y} have revealed that the Golenetskill $E_{\rm p}$-$F$ relation shows three main types of behavior: a non-monotonic relation (containing positive and negative power-law segments, with a distinct break typically at the peak flux), a monotonic relation (described by a single power law), and no clear trend.  
The time-resolved $E_{\rm p}$-$F$ in GRB 131231A (our case) shows a tight positive-monotonic correlation for both the rising and decaying wings in the log-log plot, but the power-law indices are quite different (Figure \ref{fig:f4}a). 
This case hence corresponds to the common type of monotonic relation.
For the rising wing, our best fit is $\rm log {\it E_{\rm p}}/ (keV)=(4.86\pm0.63)+(0.45\pm0.12)\times log{\it F}/(erg \ cm^{-2} s^{-1}),$\footnote{All error bars are given at the 95\% (2$\sigma$) confidence level.} with the number of data points $N$=11, the Spearman's rank correlation coefficient $R$=0.63, and a chance probability $p< 10^{-4}$; while for the decaying wing $\rm log {\it E_{\rm p}}/ (keV)=(5.09\pm0.40)+(0.58\pm0.07)\times log{\it F}/(erg \ cm^{-2} s^{-1})$  ($N$=15, $R$=0.83, $p< 10^{-4}$). The slope $S_{\rm d}$=0.58$\pm$0.07 for the decaying wing is greater than that for the rising wing $S_{\rm r}$=0.45$\pm$0.12. 
The results of our linear regression analysis for parameter relations are reported in Table \ref{tab:table2}.

The best fit to the time-resolved $\alpha$-$F$ relation \cite[e.g.,][]{2018arXiv181007313Y, 2019MNRAS.484.1912R} gives $\rm \alpha=(1.16\pm0.23)+(0.42\pm0.04)\times log{\it F}/(erg \ cm^{-2} s^{-1})$ ($N$=11, $R$=0.92, $p< 10^{-4}$) for the rising wing and  $\rm \alpha=(1.75\pm0.37)+(0.54\pm0.07)\times log{\it F}/(erg \ cm^{-2} s^{-1})$ ($N$=15, $R$=0.83, $p< 10^{-4}$) for the decaying wing.
Thus, the $\alpha$-$F$ relation (Figure \ref{fig:f4}b) for GRB 131231A is similar to the $E_{\rm p}$-$F$ relation, showing a monotonic positive relation. 

Another important relation, i.e., the $E_{\rm p}$-$\alpha$ relation, has been studied in prior works \citep[e.g.,][]{2000ApJ...543..722L, 2002ApJ...565..182L, 2006ApJS..166..298K, 2015MNRAS.451.1511B, 2015ApJ...802..132C}.
The relation for single pulses \citep[e.g.,][]{2018arXiv181007313Y} shows three main types of behaviors, similar to those of the $E_{\rm p}$-$F$ relation. 
For GRB 131231A, the best linear fit to the time-resolved $E_{\rm p}$-$\alpha$ relation gives $\rm log {\it E_{\rm p}}/ (keV)=(3.38\pm0.35)+(0.90\pm0.32)\times \alpha $ ($N$=11, $R$=0.46, $p< 10^{-4}$) for the rising wing, while $ \rm log {\it E_{\rm p}}/ (keV)=(2.92\pm0.22)+(0.84\pm0.19)\times \alpha$ ($N$=15, $R$=0.60, $p< 10^{-4}$) for the decaying wing.
With the similar relationship in the rising and the decaying wings, a tight monotonic positive relation is obtained (see Figure \ref{fig:f5}). 

Since GRB 131231A has a known redshift, we can calculate the isotropic luminosity for all the time-resolved spectra. After correcting $E_{\rm p}$ to the burst rest-frame, we show the time-resolved $E^{\rm rest}_{\rm p}$-$L_{\gamma,\rm iso}$ relation in Figure \ref{fig:f6}. 
For comparison, we also plot this relation for the GRBs reported in \cite{2010PASJ...62.1495Y}, with the time-integrated $E^{\rm rest}_{\rm p}$ and the peak isotropic luminosity $L_{\gamma,\rm iso}$ of an individual burst (see also, \citealt{2010A&A...511A..43G, 2012ApJ...754..138F, 2012ApJ...756..112L}).
The $E^{\rm rest}_{\rm p}$-$L_{\gamma,\rm iso}$ relation for the time-resolved spectra of GRB 131231A is consistent with that for the time-integrated spectra of the Yonetoku sample (101 bursts).  
More interestingly, compared with the decaying phase, the relation in the rising phase is much more compatible with that of the Yonetoku sample (Figure \ref{fig:f6}). 
Our best linear fit to the time-resolved $E^{\rm rest}_{\rm p}$-$L_{\gamma,\rm iso}$ relation gives  $\rm log {\it E^{\rm rest}_{\rm p}}/ (keV)=(-21.09\pm6.01)+(0.45\pm0.12)\times log {\it L_{\gamma,\rm iso}}/(erg \ s^{-1})$ ($N$=11, $R$=0.63, $p< 10^{-4}$) for the rising wing, while $ \rm log {\it E^{\rm rest}_{\rm p}}/ (keV)=(-28.09\pm3.78)+(0.58\pm0.07)\times log {\it L_{\gamma,\rm iso}}/(erg \ s^{-1})$ ($N$=15, $R$=0.83, $p< 10^{-4}$) for the decaying wing.
The Yonetoku's sample gives $\rm log {\it E^{\rm rest}_{\rm p}}/ (keV)=(-18.24\pm1.80)+(0.39\pm0.03)\times log {\it L_{\gamma,\rm iso}}/(erg \ s^{-1})$ ($N$=101, $R$=0.56, $p< 10^{-4}$).

\section{Physical Implications} \label{sec:models}

In short, several noticeable features of GRB 131231A can be summarized as: 
(i) the prompt emission generally displays a single large peak profile;
(ii) $\alpha$ evolution does not exceed the synchrotron limits (from -3/2 to -2/3) in all slices across the pulse;
(iii) both the $E_{\rm p}$ and the $\alpha$ evolution exhibit ``flux-tracking'' patterns across the pulse (``double-tracking''); 
(iv) the parameter relations, i.e., $E_{\rm p}$-$F$, $\alpha$-$F$, and $E_{\rm p}$-$\alpha$, along with the analogous Yonetoku $E^{\rm rest}_{\rm p}$-$L_{\gamma,\rm iso}$ relation, exhibit a strong-positive-monotonous correlation, both in the rising and the decaying wings of the pulse. 
All of these facts suggest that the spectral evolution in GRB 131231A is very interesting, which calls for physical interpretations.

Whether the GRB prompt emission is produced by synchrotron radiation or quasi-thermal emission from the photosphere \cite[e.g.,][]{2014IJMPD..2330003V, 2017IJMPD..2630018P} has been discussed and debated for a long time. Synchrotron radiation is expected in models like the internal shock model \citep{1994ApJ...430L..93R, 2011A&A...526A.110D} or the abrupt magnetic dissipation models \citep{2011ApJ...726...90Z, 2015ApJ...805..163D, 2019ApJ...882..184L}.
On the other hand, the photosphere models can be grouped into dissipative \citep{2005ApJ...628..847R,2006ApJ...652..482P, 2008A&A...480..305G, 2009ApJ...703.1044B, 2009ApJ...700L.141L, 2010PThPh.124..667I, 2011MNRAS.415.3693R, 2011MNRAS.415.1663T, 2013MNRAS.436L..54A} and nondissipative \citep{2008ApJ...682..463P, 2011ApJ...737...68B, 2011ApJ...732...49P, 2013ApJ...767..139B, 2013MNRAS.428.2430L, 2013ApJ...772...11R, 2014NewA...27...30R, 2014ApJ...785..112D, 2018ApJ...860...72M} models. 

In the following sections, we discuss whether the coexistence of ``flux-tracking'' patterns for both $\alpha$ and $E_{\rm p}$ can be understood within the frameworks of the synchrotron and photosphere models, respectively.

\subsection{Synchrotron models}
 
We consider two possible dissipation scenarios: the first scenario invokes small-radii internal shocks, with the radius defined by $R_{\rm IS} \simeq \Gamma^2 c \Delta t$, where $\Delta t$ is the observed rapid variability time scale. The second scenario invokes a large-radius internal magnetic dissipation radius, with the emission radius defined by $R_{\rm ICMART} \simeq \Gamma^{2} c t_{\rm pulse}$, where $t_{\rm pulse}$ is the duration of the entire pulse (usually the rising phase), e.g. the Internal-Collision-induced MAgnetic Reconnection and Turbulence (ICMART) model \citep{2011ApJ...726...90Z}. In this second scenario, the rapid variability time scale is related to the mini-jets associated with local magnetic reconnection sites in the ejecta \citep{2014ApJ...782...92Z}. The former model invokes multiple emission sites, i.e. emission from many internal shocks contribute to the observed emission, while the latter model invokes one emitter, which continuously radiate as it streams away from the engine. So it can be regarded as a one-zone model.

In the framework of the synchrotron model, the peak energy can be written as $E_{\rm p} \propto L^{1/2} \gamma^{2}_{\rm e,ch} R^{-1}(1+z)^{-1}$, where $L$ is the `wind' luminosity of the ejecta, $\gamma_{\rm e, ch}$ is the typical electron Lorentz factor in the emission region, $R$ is the emission radius, and $z$ is the redshift \citep{2002ApJ...581.1236Z}. If other parameters are similar to each other, one naturally has $E_{\rm p} \propto L^{1/2}$, and hence, a tracking behavior. This is more straightforward for the small-radii internal shock model. For the ICMART model, since the emitter is initially at a smaller radius where the magnetic field is stronger, the $E_{\rm p}$ evolution likely shows a hard-to-soft evolution \citep{2011ApJ...726...90Z,2014NatPh..10..351U, 2016ApJ...825...97U}. On the other hand, considering other factors such as bulk acceleration, \cite{2018ApJ...869..100U} showed that the one-zone synchrotron model can produce both hart-to-soft and flux-tracking patterns depending on parameters. The $E_{\rm p}$ flux tracking, therefore, can be made consistent with both synchrotron models.

The clue to differentiate between the models comes from the $\alpha$ tracking behavior, as observed in GRB 131231A. 
In the rising phase, $\alpha$ gets harder, which suggests that the emitting electrons are evolving from the fast cooling regime to the slow cooling regime. 
Invoking many emission regions (like in the small-radii internal shock model) to satisfy this constrain required very contrived coincidence. On the other hand, the large-radius one-zone model can do this naturally, as shown in \cite{2014NatPh..10..351U}; as the emission region moves away from the central engine, the magnetic field in the emission region naturally decays with time.
This would cause accelerated electrons to experience a history of different degrees of cooling at different times, e.g. from fast cooling to slow cooling. The resulting photon spectrum should also experience an evolution from fast-cooling-like to slow-cooling-like.
Another way to harden $\alpha$ is to introduce the transition from  synchrotron cooling to synchrotron self-Compton cooling in the Klein-Nishina regime for the electrons \citep{2009A&A...498..677B, 2011A&A...526A.110D, 2018ApJS..234....3G}. Since $E_{\rm p}$ increases in the rising phase, the characteristic Lorentz factor $\gamma_{\rm e,ch}$ of emitting electrons should be increasing with $R$ when the magnetic field is decaying. 
The increase of $\gamma_{\rm e, ch}$ is consistent with the particle-in-cell simulations \cite[e.g., ][]{2017ApJ...843L..27W, 2018MNRAS.481.5687P}.
Such an increase would enhance synchrotron self-Compton cooling of electrons (see Equation 27 in \citealt{2018ApJS..234....3G}). On the other hand, the increasing flux intensity\footnote{When the magnetic field is decreasing, the flux density could increase if the 
characteristic Lorentz factor $\gamma_{\rm e,ch}$ or the injection rate of electrons is increasing.} indicates that the ratio of the radiation energy density to the magnetic energy density is rising, which also supports that the synchrotron self-Compton cooling for electrons is getting more significant.
These effects together will make the spectrum of cooling electrons hard.
Therefore, both $E_{\rm p}$ and $\alpha$ tracking the flux intensity is naturally interpreted within the one-zone synchrotron model during the rising phase.

In the decaying wing, $\alpha$ gets observationally softer. This is also understandable within this theoretical framework. The decay phase of a broad pulse is likely controlled by the so-called ``curvature effect'' \citep{2000ApJ...541L..51K, 2015ApJ...808...33U, 2016ApJ...825...97U}, which predicts an $\hat\alpha=2+\hat\beta$ closure relation (in the convention of $F_\nu \propto t^{-\hat\alpha} \nu^{-\hat\beta}$) if a proper time zero-point is chosen \citep{2006ApJ...642..354Z}\footnote{Another way to interpret the decaying phase of a pulse is to assume that the accelerated electrons have a progressively lower minimum energy at the shock front \citep{2011A&A...526A.110D}. However, there is no predictable closure relation to test this model.}. In Figure \ref{fig:f7}, we test this closure relation, which suggests that the relation is roughly satisfied. With this interpretation, both $E_{\rm p}$ and $\alpha$ are expected to track the flux. This is because when the dissipation process ceases abruptly, the observer would observe emission from progressively higher latitudes, which corresponds to an earlier emission time. One would then observe a reversely softening spectrum during the decay phase. Due to the progressively lower Doppler factor at higher latitudes, $E_{\rm p}$ also decays with time during the decaying phase.

As a result, the observed ``double-tracking'' behavior can be well interpreted within the one-zone synchrotron model.

\subsection{Photosphere models}

The photosphere models \citep{2005ApJ...628..847R,2006ApJ...652..482P, 2008A&A...480..305G, 2008ApJ...682..463P,  2009ApJ...703.1044B, 2011ApJ...737...68B, 2009ApJ...700L.141L, 2010PThPh.124..667I, 2011ApJ...732...49P, 2011MNRAS.415.3693R, 2011MNRAS.415.1663T, 2013MNRAS.436L..54A, 2013ApJ...767..139B, 2013MNRAS.428.2430L, 2013ApJ...772...11R, 2014NewA...27...30R, 2014ApJ...785..112D, 2018ApJ...860...72M} invoke an even smaller emission radius than the small-radii internal shock model, which interprets the broad pulse as observed in GRB 131231A as a manifestation of the history of central engine activity. Since usually the luminosity is positively correlated to the temperature, the photosphere model usually predicts an $E_{\rm p}$-flux-tracking behavior \citep{2014ApJ...785..112D}, even though in certain structured jet geometry, a reversed (hard-to-soft evolution) pattern may be also produced \citep{2019ApJ...882...26M}.

The difficulty is to produce the observed $\alpha$ and its tracking behavior. There are several issues. First, the photosphere models, even with the temporal and spatial superposition effects considered, predict a much harder value ($\alpha \sim + 0.4$) than observed \citep{2014ApJ...785..112D}. In order to reproduce a typical $\alpha \sim -1$, a special jet structure needs to be introduced \citep{2013MNRAS.428.2430L}. Second, similar to the small-radii internal shock model, in order to reproduce the well-observed $\alpha$-tracking behavior, very contrived conditions from the central engine (the power from the engine as well as the jet structure from the engine) are needed. Both an overall soft $\alpha$ value and the nice double-tracking behavior of GRB 131231A disfavor a photosphere interpretation of this burst.

\section{Discussion and Conclusion} \label{sec:Conclusion}

It is useful to compare our finding with several previous papers. \cite{2009A&A...505..569B} selected a sample of 18 GRBs that have at least two time-resolved spectra of the precursor, from 51 BATSE bursts with the precursor presented in \cite{2006ApJS..166..298K}. They investigated the relationship of the spectral feature between the precursor and the main GRB episodes. Out of 18 bursts, they found 1 case (GRB 930201) whose both photon spectral index $\alpha$ and $E_{\rm p}$ follow a strong soft-to-hard evolution in the rising phase of the precursor and vice versa in the descending part (see Figure 4 in \citealt{2009A&A...505..569B}). This is similar to our case, but the trend is not as clear as our case. The rising part of the $E_{\rm p}$ evolution for GRB 930201 does not exhibit an ``ideal'' flux-tracking behavior due to two reasons: first, in their time-resolved spectral analysis, the number of time bins (only four) is limited; second, the first time bin obviously deviates from the flux-tracking behavior. 
We also compared the $E_{\rm p}$-$\alpha$ relation in our case with three other single pulse bursts studied by \cite{2012ApJ...756..112L}, in which $E_{\rm p}$ also exhibits the flux-tracking behavior. No clear relationship is found for those three bursts and much harder $\alpha$ values are derived (see Figure \ref{fig:f5}). Recently, \cite{2018arXiv181007313Y} systematically studies a complete catalog of the spectral evolution of 38 single pulses from 37 \textit{Fermi} GRBs with a fully Bayesian approach and found that the $\alpha$ evolution does not show a strong general trend.

In this paper, we report both $E_{\rm p}$ and $\alpha$ evolutions of GRB 131231A that show ``flux-tracking'' characteristics simultaneously (``double-tracking'') across its entire single pulse. All the parameter relations, i.e., $E_{\rm p}$-$F$, $\alpha$-$F$, and $E_{\rm p}$-$\alpha$ relations, along with the Yonetoku $E^{\rm rest}_{\rm p}$-$L_{\gamma,\rm iso}$ relation, exhibit strong-positive-monotonous correlations, both in the rising and the decaying wings. Such ``double-tracking'' features are rarely observed within single pulse bursts, and this is the first ideal case showing that the $E_{\rm p}$, as well as the $\alpha$, simultaneously track the flux.
We then discuss how these unique characteristics of spectral evolution may be interpreted within the frameworks of both the synchrotron and photosphere models. We find that the coexistence of the flux-tracking behaviors for $E_{\rm p}$ and $\alpha$ can be naturally interpreted with the one-zone synchrotron emission model, with the emission region far from the central engine. It disfavors the photosphere origin of emission from this burst.  
We expect that similar features may exist in more bursts. In fact, dedicated searches of these features in a larger sample (D. Tak et al. 2019, in preparation) indeed revealed similar features in a larger sample independently.

In conclusion, our analysis suggests that the spectral evolution in GRB 131231A is the first ideal case showing that the $E_{\rm p}$ and $\alpha$ both simultaneously track the flux so far. Considering other features--- for instance, single large peak pulse, $\alpha$ evolution does not exceed the synchrotron limit ($\alpha$= -2/3) in all slices across the pulse--- altogether such distinct features have never been identified simultaneously in a single GRB in the previous observations.

\acknowledgments
We appreciate to the referee for the constructive report.
We acknowledge the use of the public data from the {\it Fermi} data archives.
The majority of this work were performed during L. L.'s visit to the Purple Mountain Observatory of Chinese Academy of Sciences, Nanjing, in the summer of 2018. 
This work is partially supported by the National Natural Science Foundation of China (grants Nos. 11673068, 11725314, and 11873030), the Key Research Program of Frontier Sciences (grant No. QYZDB-SSW-SYS005), the Strategic Priority Research Program ``Multi-waveband gravitational wave Universe'' (grant No. XDB23000000) of the Chinese Academy of Sciences, and the ``333 Project'' of Jiangsu Province. 
J.J.G. acknowledges the support from the National Postdoctoral Program for Innovative Talents (grant No. BX201700115), the China Postdoctoral Science Foundation funded project (grant No. 2017M620199).

\vspace{5mm}
\facilities{{\it Fermi}/GBM}
\software{RMFIT, XSPEC \citep{1996ASPC..101...17A}, 3ML\citep{2015arXiv150708343V}, and Matplotlib\citep{2007CSE.....9...90H}} 
\bibliography{../../MyBibFiles/MyBibFile.bib}

\begin{deluxetable}{ccccccc}
\tablecaption{Results of the Time-resolved Spectral Fits of GRB 131231A \label{tab:table1}}
\tablehead{
\colhead{$t_{1}$$\sim$$t_{2}$\tablenotemark{a}}
& \colhead{S/N} 
&\colhead{$\alpha$}
&\colhead{$\beta$}
&\colhead{$E_{\rm p}$}
&\colhead{C-stat/dof}
&\colhead{Red.C-stat}\\
(s)&&&&(keV)
}
\colnumbers
\startdata
0.000$\sim$2.886&14.58&-1.51$\pm$0.13&-2.12$\pm$0.11&53.74$\pm$8.68&401.40/314&1.28\\
2.886$\sim$12.683&38.83&-1.36$\pm$0.11&-2.17$\pm$0.15&124.50$\pm$29.30&573.88/314&1.83\\
12.683$\sim$15.056&35.63&-1.18$\pm$0.11&-2.38$\pm$0.38&184.70$\pm$40.50&406.14/314&1.29\\
15.056$\sim$15.437&22.03&-1.28$\pm$0.07&unconstrained&495.70$\pm$126.35&404.12/314&1.29\\
15.437$\sim$16.165&48.56&-1.04$\pm$0.05&unconstrained&471.90$\pm$53.90&338.15/314&1.08\\
16.165$\sim$17.968&68.45&-1.05$\pm$0.05&-2.38$\pm$0.24&317.50$\pm$38.00&342.01/314&1.09\\
17.968$\sim$19.966&106.06&-0.92$\pm$0.04&-2.34$\pm$0.11&275.10$\pm$19.40&331.56/314&1.06\\
19.966$\sim$21.278&101.66&-0.92$\pm$0.04&-2.60$\pm$0.20&282.90$\pm$18.60&341.45/314&1.09\\
21.278$\sim$21.597&59.67&-0.87$\pm$0.08&-2.44$\pm$0.23&277.50$\pm$33.60&298.70/314&0.95\\
21.597$\sim$22.221&97.06&-0.96$\pm$0.03&unconstrained&473.00$\pm$25.60&315.04/314&1.00\\
22.221$\sim$23.375&165.80&-0.85$\pm$0.03&-2.67$\pm$0.13&346.10$\pm$13.80&379.86/314&1.21\\
23.375$\sim$24.659&155.46&-0.87$\pm$0.03&-2.77$\pm$0.15&242.40$\pm$10.10&351.51/314&1.12\\
24.659$\sim$25.263&116.19&-0.81$\pm$0.05&-2.49$\pm$0.12&239.70$\pm$14.70&404.67/314&1.29\\
25.263$\sim$26.353&145.40&-0.91$\pm$0.04&-2.40$\pm$0.08&197.90$\pm$11.30&351.35/314&1.12\\
26.353$\sim$27.721&138.03&-0.98$\pm$0.05&-2.32$\pm$0.07&140.40$\pm$9.67&389.81/314&1.24\\
27.721$\sim$29.192&129.90&-0.98$\pm$0.07&-2.41$\pm$0.07&96.57$\pm$6.11&340.68/314&1.08\\
29.192$\sim$31.152&138.79&-1.06$\pm$0.08&-2.37$\pm$0.06&73.37$\pm$4.51&293.85/314&0.94\\
31.152$\sim$32.195&117.85&-1.20$\pm$0.07&-2.76$\pm$0.16&86.39$\pm$5.10&403.77/314&1.29\\
32.195$\sim$33.320&108.76&-1.21$\pm$0.07&-3.02$\pm$0.24&74.84$\pm$3.82&365.74/314&1.16\\
33.320$\sim$35.454&117.18&-1.16$\pm$0.08&-2.65$\pm$0.10&57.98$\pm$2.93&441.12/314&1.40\\
35.454$\sim$37.245&83.90&-1.10$\pm$0.19&-2.30$\pm$0.06&45.62$\pm$4.87&365.87/314&1.17\\
37.245$\sim$38.427&86.84&-1.37$\pm$0.07&-2.74$\pm$0.26&87.99$\pm$7.69&347.67/314&1.11\\
38.427$\sim$40.354&86.40&-1.42$\pm$0.08&-2.59$\pm$0.18&74.13$\pm$6.70&401.45/314&1.29\\
40.354$\sim$41.834&56.08&-1.31$\pm$0.18&-2.37$\pm$0.13&55.82$\pm$8.22&312.84/314&1.00\\
41.834$\sim$46.031&67.02&-1.57$\pm$0.09&-2.72$\pm$0.29&54.80$\pm$4.75&366.55/314&1.17\\
46.031$\sim$50.000&40.23&-1.48$\pm$0.19&-2.63$\pm$0.24&39.85$\pm$4.33&410.26/314&1.31\\
\enddata
\tablenotetext{a}{Time intervals.}
\end{deluxetable}

\clearpage
\begin{deluxetable*}{ccccccccc}
\tablewidth{0pt}
\tablecaption{Results of our Linear Regression Analysis for Parameter Relations}\label{tab:table2}
\tablenum{2}
\tablehead{
\colhead{Relation}&
\colhead{Phase}&
\colhead{Expression}&
\colhead{$R$}&
\colhead{$p$}&
}
\startdata
$E_{\rm p}$\tablenotemark{a}-$F$\tablenotemark{b}&Rising&log$E_{\rm p}$=(4.86$\pm$0.63)+(0.45$\pm$0.12)$\times$log$F$&0.63&$<$10$^{-4}$\\
$E_{\rm p}$-$F$&Decaying&log$E_{\rm p}$=(5.09$\pm$0.40)+(0.58$\pm$0.07)$\times$log$F$&0.83&$<$10$^{-4}$\\
$\alpha$-$F$&Rising&$\alpha$=(1.16$\pm$0.23)+(0.42$\pm$0.04)$\times$log$F$&0.92&$<$10$^{-4}$\\
$\alpha$-$F$&Decaying&$\alpha$=(1.75$\pm$0.37)+(0.54$\pm$0.07)$\times$log$F$&0.83&$<$10$^{-4}$\\
$E_{\rm p}$-$\alpha$&Rising&log$E_{\rm p}$=(3.38$\pm$0.35)+(0.90$\pm$0.32)$\times$$\alpha$&0.46&$<$10$^{-4}$\\
$E_{\rm p}$-$\alpha$&Decaying&log$E_{\rm p}$=(2.92$\pm$0.22)+(0.84$\pm$0.19)$\times$$\alpha$&0.60&$<$10$^{-4}$\\
$E^{\rm rest}_{\rm p}$\tablenotemark{a}-$L_{\gamma,\rm iso}$\tablenotemark{c}&Rising&log$E_{\rm p}$=(-21.09$\pm$6.01)+(0.45$\pm$0.12)$\times$log$L_{\gamma,\rm iso}$&0.63&$<$10$^{-4}$\\
$E^{\rm rest}_{\rm p}$-$L_{\gamma,\rm iso}$&Decaying&log$E_{\rm p}$=(-28.09$\pm$3.78)+(0.58$\pm$0.07)$\times$log$L_{\gamma,\rm iso}$&0.83&$<$10$^{-4}$\\
\enddata
\tablenotetext{a}{In units of keV.}
\tablenotetext{b}{In units of erg cm$^{-2}$ s$^{-1}$.}
\tablenotetext{c}{In units of erg s$^{-1}$.}
\end{deluxetable*}

\clearpage
\begin{figure*}
\centering
\plotone{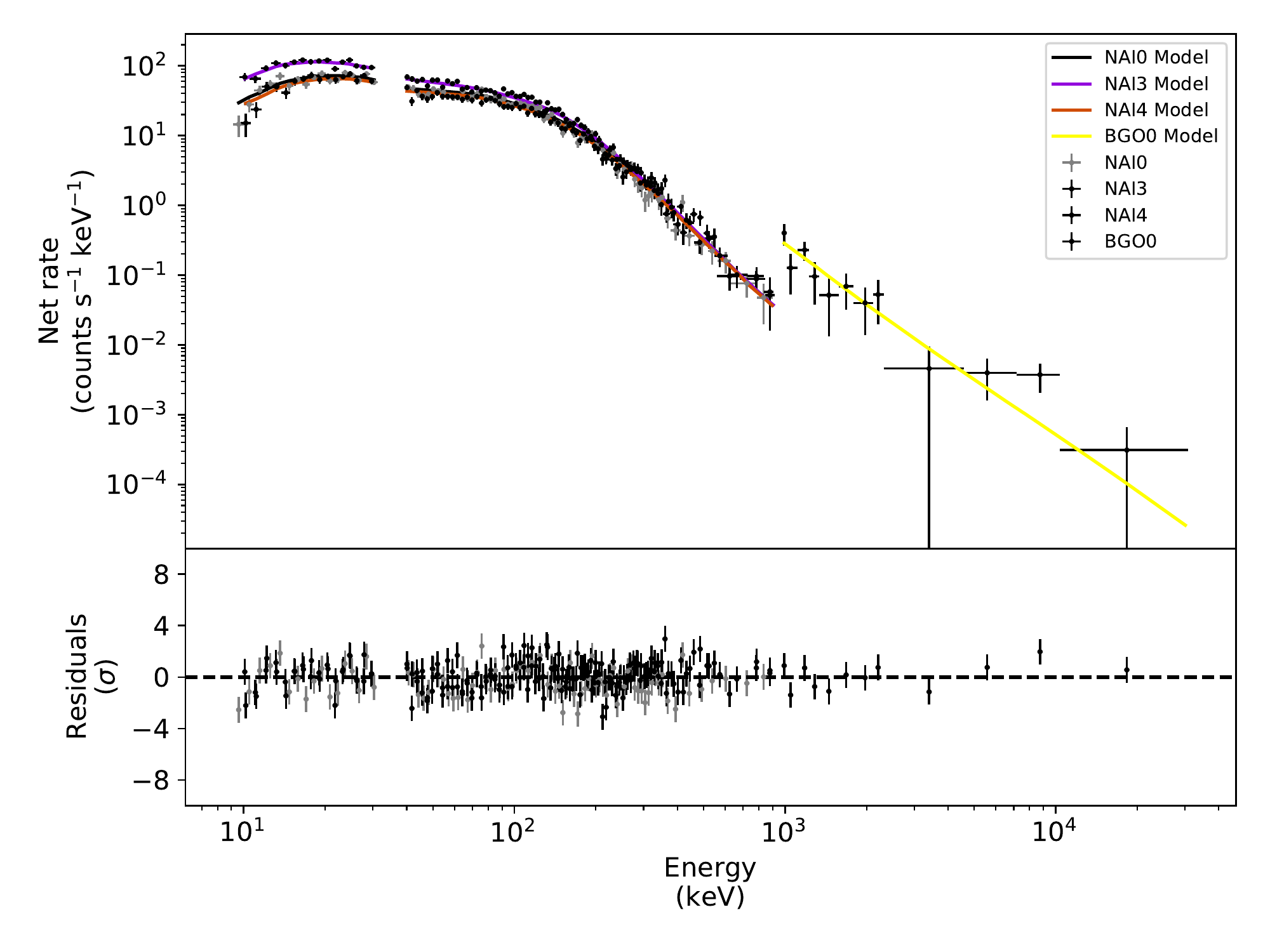}
\caption{Example of count spectral fit results using the brightest (highest S/N) time bin (22.221$\sim$23.375).}
\end{figure*}\label{fig:f1}

\clearpage
\begin{figure*}
\gridline{\fig{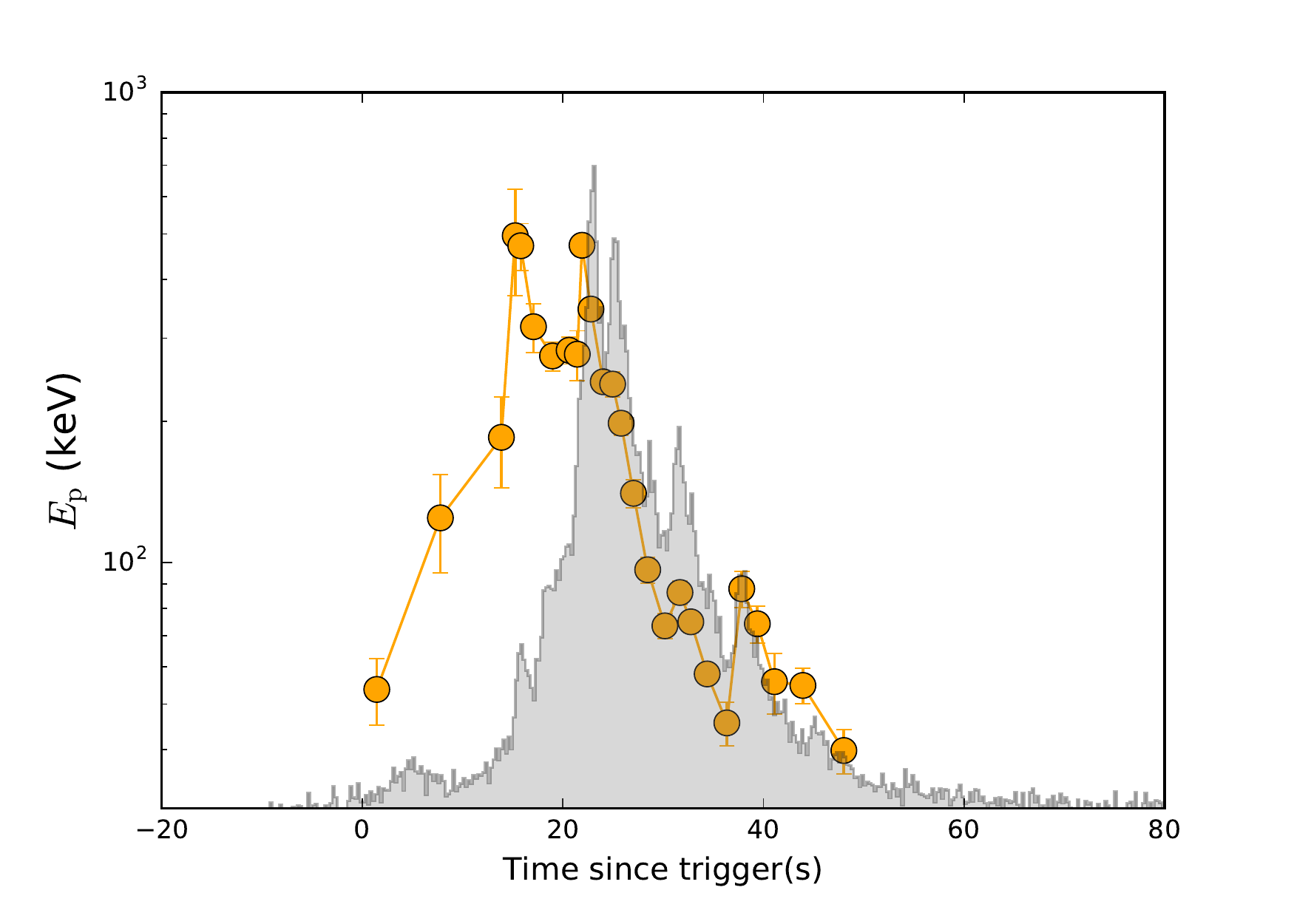}{0.5\textwidth}{}
          \fig{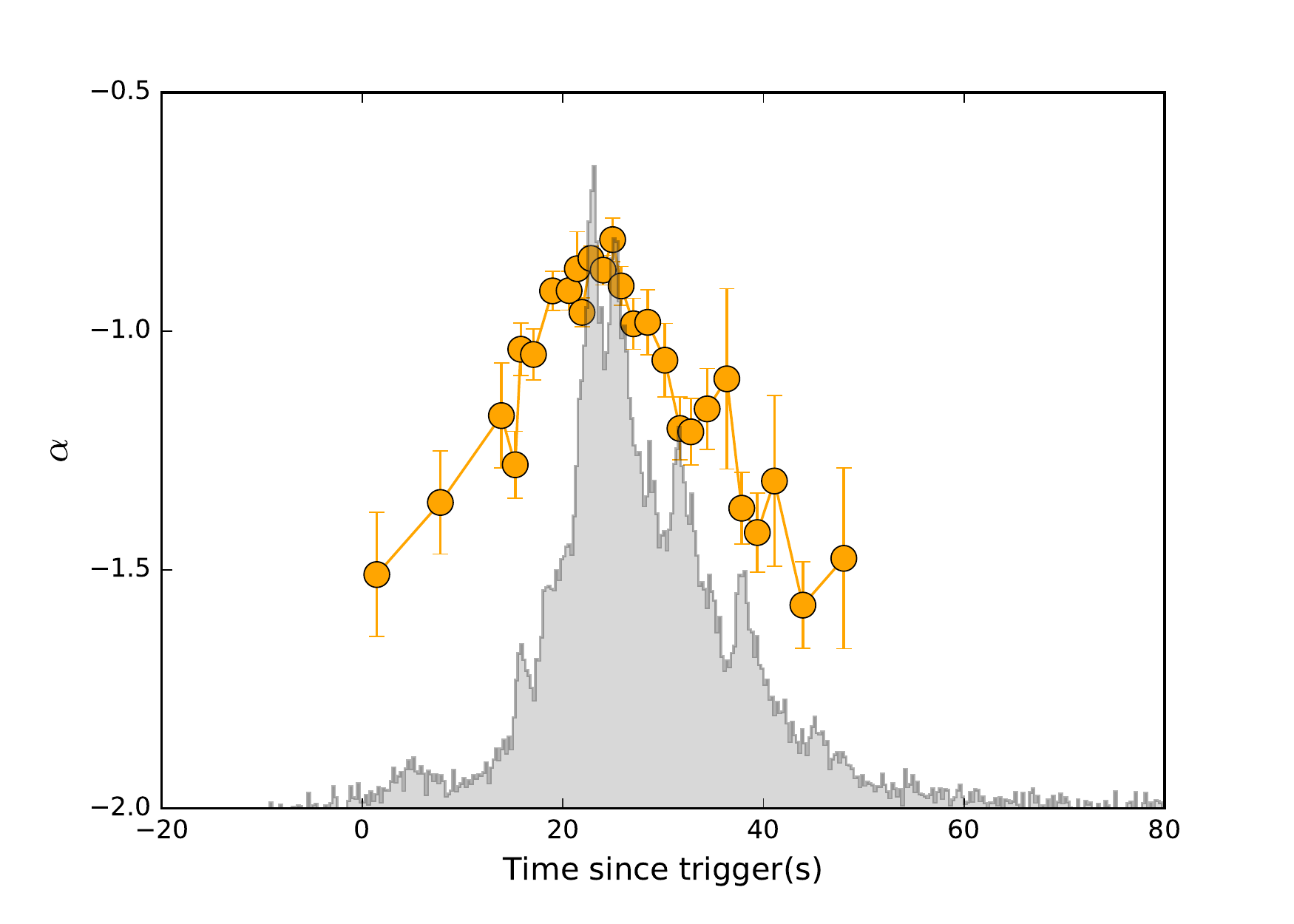}{0.5\textwidth}{}
          }
\caption{Temporal evolution of the $E_{\rm p}$ (marked with orange color in the left panel) and $\alpha$ (marked with orange color in the right panel) of GRB 131231A, and the GRB light curve is overlaid in grey.}
\end{figure*}\label{fig:f2}

\clearpage
\begin{figure*}
\gridline{\fig{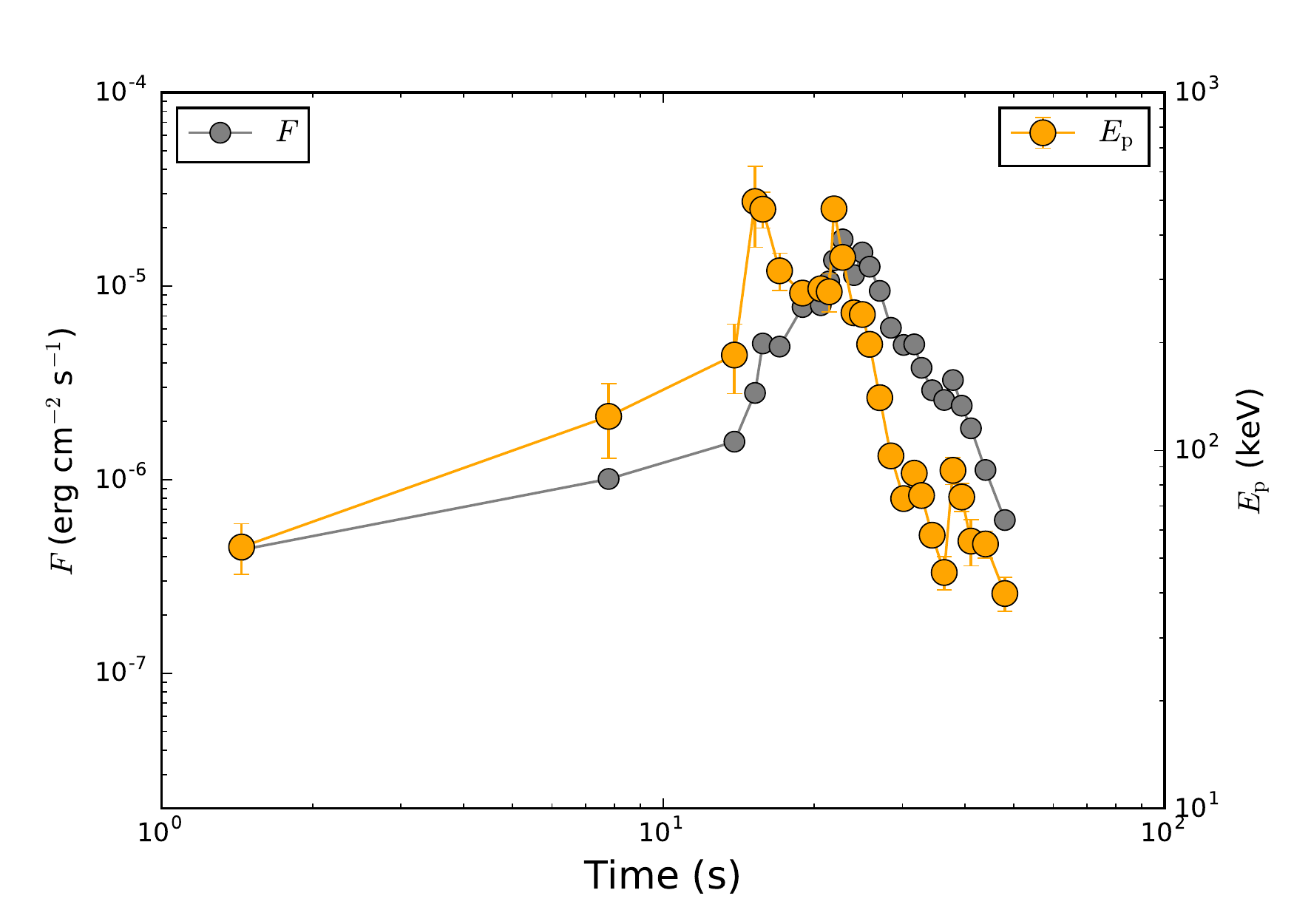}{0.5\textwidth}{}
          \fig{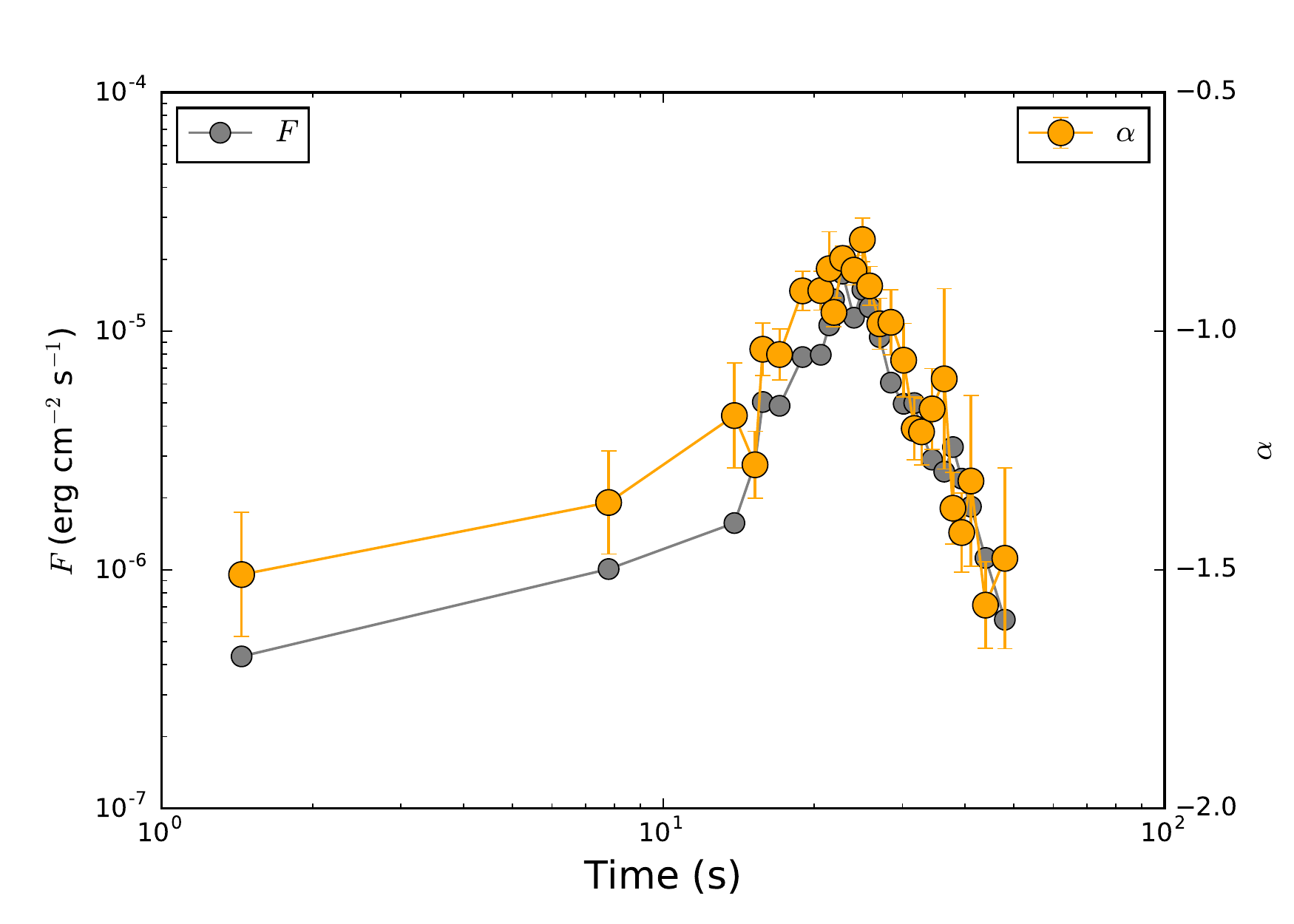}{0.5\textwidth}{}
          }
\caption{Temporal evolution of the energy flux (the left-hand y-axis), along with $E_{\rm p}$ (left panel) and $\alpha$ (right panel). The symbols are the same as in Figure \ref{fig:f2}.}
\end{figure*}\label{fig:f3}

\clearpage
\begin{figure*}
\gridline{\fig{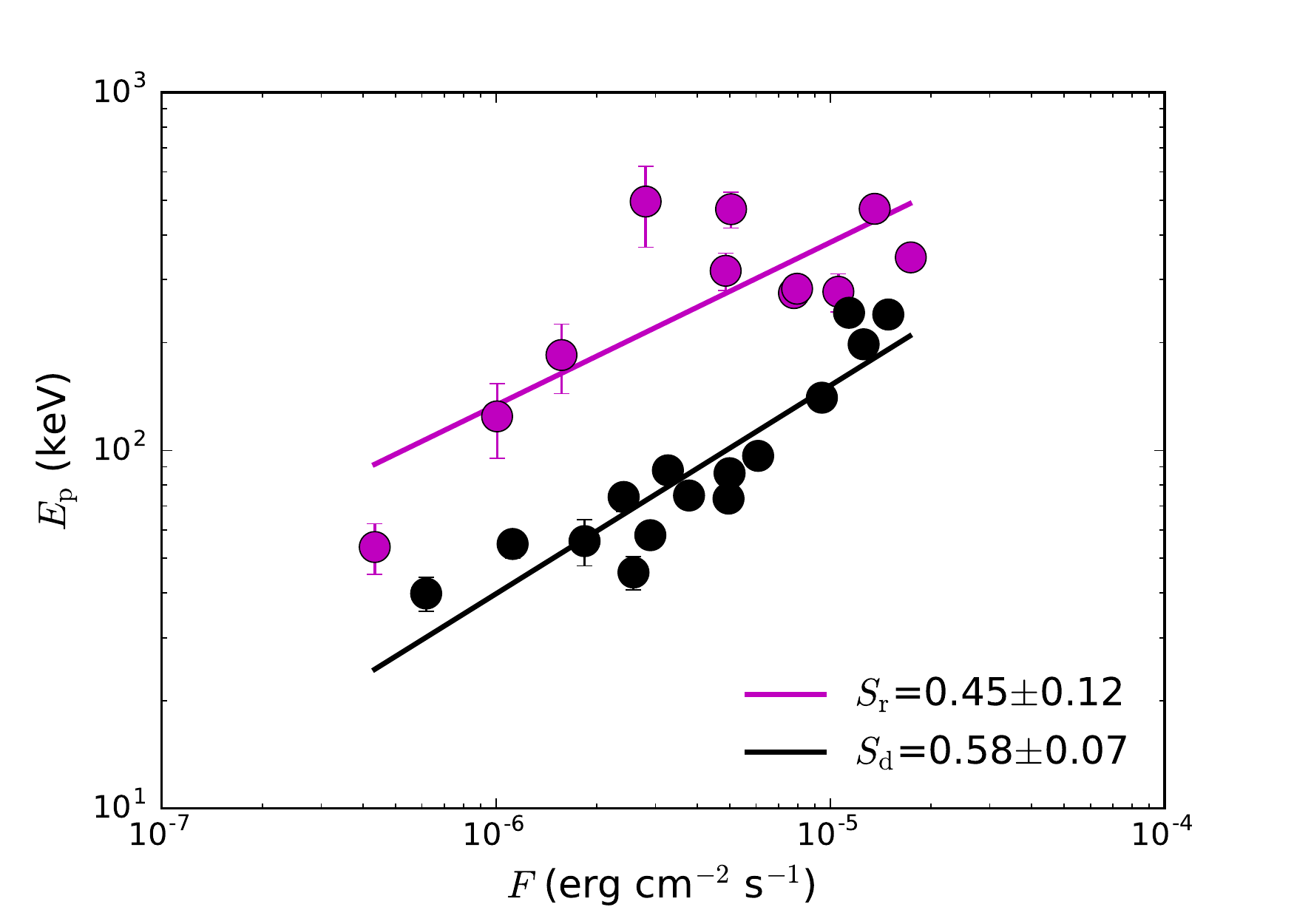}{0.5\textwidth}{}
          \fig{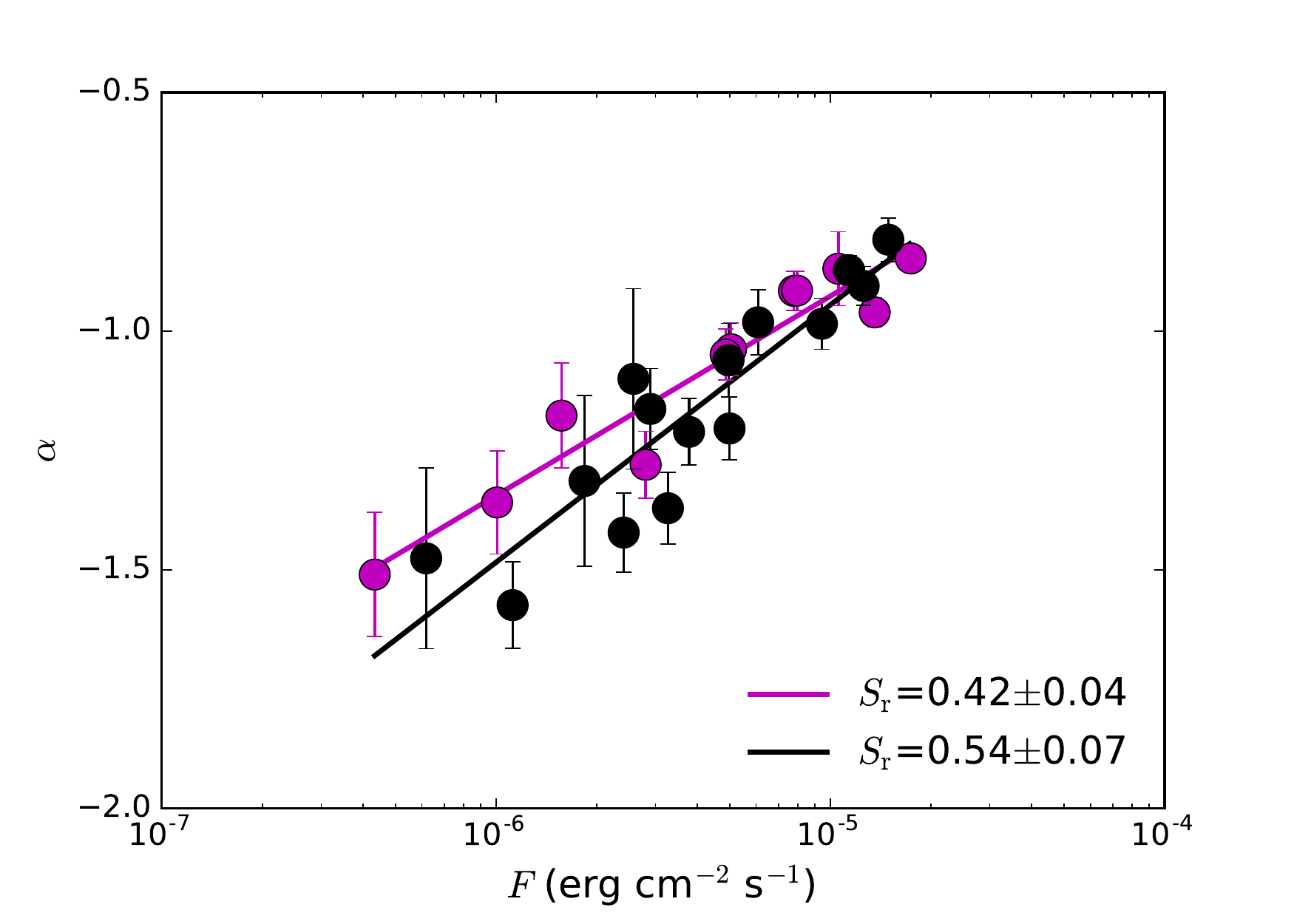}{0.5\textwidth}{}
          }
\caption{$E_{\rm p}$ (left panel), and $\alpha$ (right panel) as a function of the flux. The purple dot and the black dot data points represent the rising and decaying wing, respectively. The solid black and solid purple lines represent the best fit to the rising and decaying wings, respectively.}
\end{figure*}\label{fig:f4}

\clearpage
\begin{figure*}
\centering
\plotone{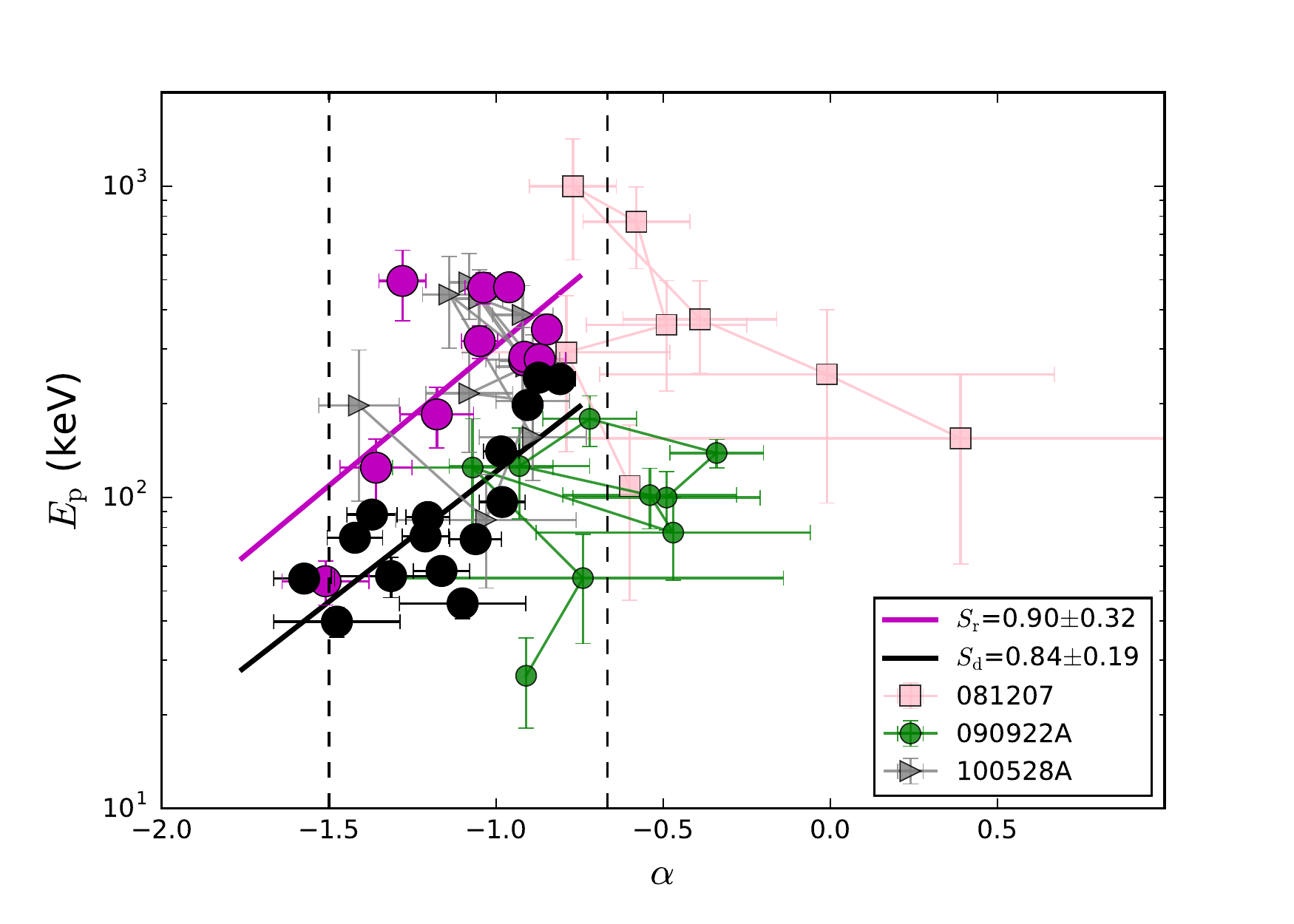}
\caption{$E_{\rm p}$-$\alpha$ relation, compared with other three single pulse bursts that exhibit a 'flux-traking' behavior for $E_{\rm p}$ evolution (GRB 081207, GRB 090922A, and GRB 100528A) studied in \cite{2012ApJ...756..112L}. Two vertical dashed lines represent the limiting values (synchrotron) of $\alpha$ =-2/3 and $\alpha$ =-3/2 for electrons in the slow and fast cooling regimes, respectively.}
\end{figure*}\label{fig:f5}

\clearpage
\begin{figure*}
\plotone{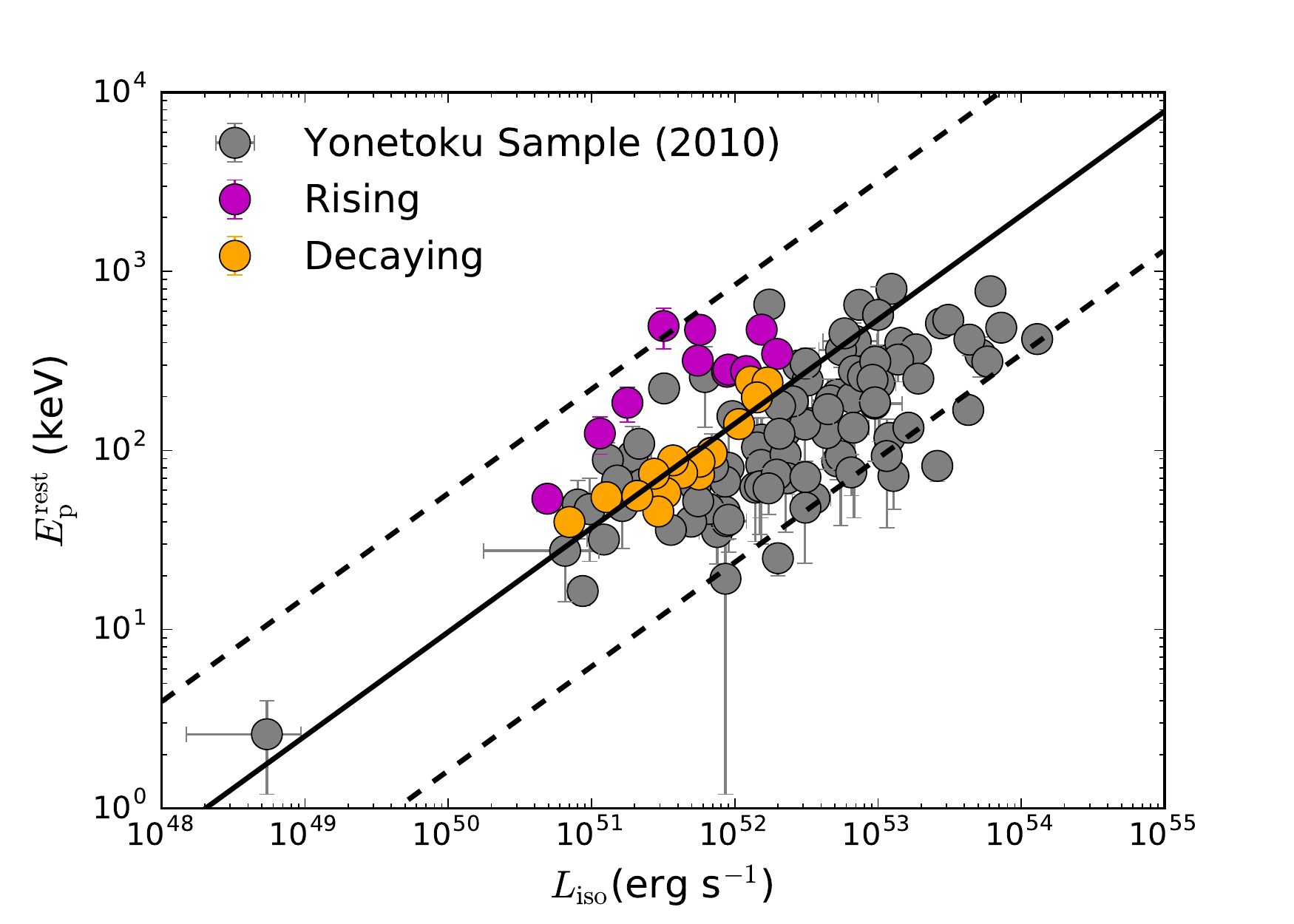}
\caption{Comparison of the time-resolved $E^{\rm rest}_{\rm p}$-$L_{\gamma,\rm iso}$ correlation for the rising (pink) and decaying (orange) phases of our case (GRB 131231A) with the time-integrated $E^{\rm rest}_{\rm p}$-$L_{\gamma,\rm iso}$ correlation for the 101 GRBs in \citealt{2010PASJ...62.1495Y} (gray filled circles). The solid line is the best fit to the time-resolved spectra of the Yonetoku sample, while the two dotted lines represent its 2$\sigma$ dispersion around the best fit.}
\end{figure*}\label{fig:f6}

\clearpage
\begin{figure*}
\plottwo{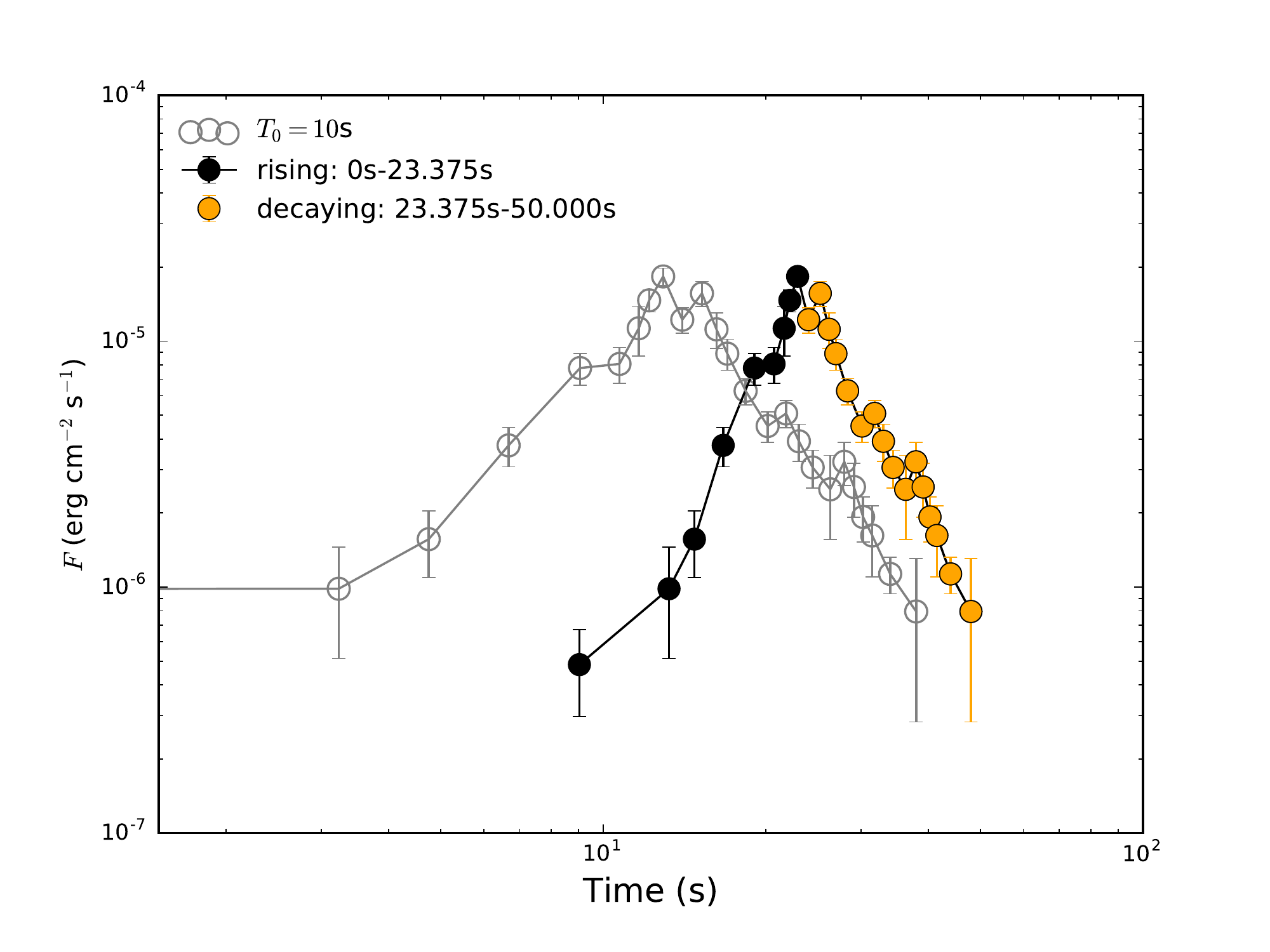}{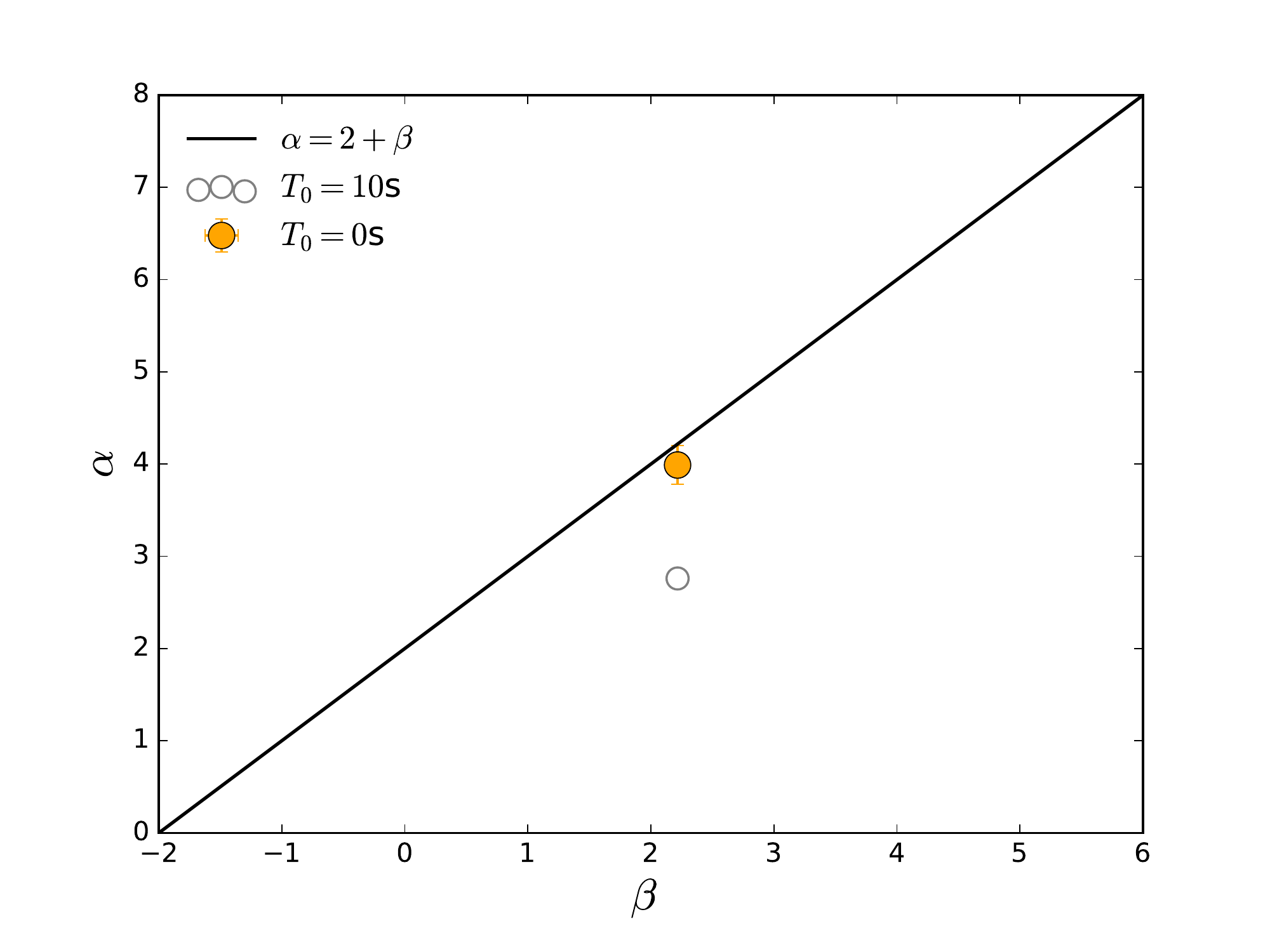}
\caption{Test of the high-latitude curvature effect in GRB 131231A. {Left panel: the {\it Fermi}-GBM light curve are marked with solid points, while 
the hollow points indicate the new {\it Fermi}-GBM light curve by shifting with $T_{0}$=10s.
Right panel: testing the closure relation of the curvature effect in the decaying wing (orange); colors are the same as in the left panel. The temporal index $\hat\alpha$ and the spectral index $\hat\beta$ satisfy a simple relation \citep{2000ApJ...541L..51K}: $\hat\alpha$= 2+$\hat\beta$ (solid line), with the convention $F^{\rm obs}_{\nu_{\rm obs}} \propto t^{-\hat\alpha}_{\rm obs} \nu^{-\hat\beta}_{\rm obs}$.}} 
\end{figure*}\label{fig:f7}

\end{document}